\documentclass{aa}
\usepackage{graphicx}
\usepackage{amsmath}
\usepackage{txfonts}
\usepackage{subfigure}

\usepackage{natbib}
\bibpunct{(}{)}{;}{a}{}{,} 

\usepackage{hyperref}

\graphicspath{{figures/}}

 \title{Shear mixing in stellar radiative zones\\I.~Effect of thermal diffusion and chemical stratification}
\author{V. Prat\inst{1,2,3} \and F. Ligni\`eres\inst{1,2}}
\institute{Universit\'e de Toulouse; UPS-OMP; IRAP; Toulouse, France \and CNRS; IRAP; 14, avenue \'Edouard Belin, F-31400 Toulouse, France \and Max-Planck Institut f\"ur Astrophysik, Karl-Schwarzschild-Str.~1, 85748, Garching bei M\"unchen, Germany}
\offprints{V. Prat}
\date{Received 17 February 2014 / Accepted 18 April 2014}
\abstract
{Turbulent transport of chemical elements in radiative zones of stars is considered in current stellar evolution codes thanks to phenomenologically derived diffusion coefficients.
Recent local numerical simulations (Prat \& Ligni\`eres 2013, \aap, 551, L3) suggest that the coefficient for radial turbulent diffusion due to radial differential rotation satisfies $D_{\rm t}\simeq0.058\kappa/Ri$, in qualitative agreement with the model of Zahn (1992, \aap, 265, 115).
However, this model does not apply (i) when differential rotation is strong with respect to stable thermal stratification or (ii) when chemical stratification has a significant dynamical effect, a situation encountered at the outer boundary of nuclear-burning convective cores.}
{We extend our numerical study to consider the effects of chemical stratification and of strong shear, and compare the results with prescriptions used in stellar evolution codes.}
{We performed local, direct numerical simulations of stably stratified, homogeneous, sheared turbulence in the Boussinesq approximation.
The regime of high thermal diffusivities, typical of stellar radiative zones, is reached thanks to the so-called small-P\'eclet-number approximation, which is an asymptotic development of the Boussinesq equations in this regime.
The dependence of the diffusion coefficient on chemical stratification was explored in this approximation.}
{Maeder's extension of Zahn's model in the strong-shear regime (Maeder 1995, \aap, 299, 84) is not supported by our results, which are better described by a model found in the geophysical literature.
As regards the effect of chemical stratification, our quantitative estimate of the diffusion coefficient as a function of the mean gradient of mean molecular weight leads to the formula $D_{\rm t}\simeq 0.45\kappa(0.12-Ri_\mu)/Ri$, which is compatible in the weak-shear regime with the model of Maeder \& Meynet (1996, \aap, 313, 140) but not with Maeder's (1997, \aap, 321, 134).}
{}

\begin{document}

\maketitle

\section{Introduction}

Mixing of chemical elements in stellar interiors is a crucial feature in stellar evolution theory.
Indeed, for main sequence stars, mixing can continuously draw hydrogen from outer layers to the core.
For a given initial mass, stars experiencing mixing have a higher averaged mean molecular weight and thus appear brighter.
The hydrogen-core-burning lifetime is increased as is the mass of the helium core. Larger helium cores modify the subsequent evolution.
It is thus essential to have a good knowledge of transport processes to build reliable stellar models.
Whereas microscopic processes, such as molecular diffusion, gravitational settling, and radiative acceleration, are relatively well known, the effects of macroscopic motions induced by rotation are still poorly understood.
Improving such models may have a strong impact in many fields of astrophysics.
This is particularly true for galaxy physics, because the properties of the stars in a galaxy provide information about the galaxy, and for planetary systems, where constraints on the stellar host help us constrain the mass and the radius of planets.

The main technique used to constrain mixing processes in stars is to determine chemical abundances on the surface of stars through measurements of the depth and the width of atomic absorption lines.
These measurements may bring out some anomalies for certain stars, that is, over- or under-abundances as compared to the standard model of stellar evolution, which describes the evolution of a spherical, non-rotating star in which the only macroscopic transport process is convection.
For massive stars, in which the CNO cycle is dominant, abundance anomalies of elements involved in this cycle, such as helium, carbon, or nitrogen, indicate that some deep mixing is occurring in the observed star (see e.g. \citealt{GiesLambert}, \citealt{Lyubimkov}, and more recently \citealt{Hunter} with the VLT-FLAMES survey).
\citet{Martins2009, Martins2013} have even found evidence for the existence of stars in which mixing is so efficient that they are quasi-homogeneous.

Another sort of anomaly is linked to the fact that light elements (lithium, beryllium, and boron) are destroyed at high temperature.
For massive stars, lithium and beryllium are totally depleted, whereas boron is only destroyed in the inner layers.
If some mixing occurs between these layers and the outer ones, the surface abundance of boron is lower than otherwise \citep{Venn, Mendel}.
For cooler stars, deep convection causes lithium and beryllium to be depleted.
It follows that these elements can only be detected in a narrow range of stellar mass.
Here again we find discrepancies between the standard model and the observed abundances, including for the Sun \citep[e.g. the lithium dip, initially observed by][]{BoesgaardTripicco}.
A source of extra mixing, such as internal gravity waves \citep{TalonCharbonnel}, is thus needed.

Stellar seismology is another powerful way to constrain transport processes within stars.
The analysis of stellar oscillations not only provides new global parameters that are useful for better determining fundamental parameters of stars, such as mass, radius, and age, but it also allows us to probe the internal properties of some stars, including the distribution of chemicals (for the Sun), \emph{via} the speed of sound \citep{ChristensenDalsgaard, AntiaBasu} and internal rotation.
\citet{Beck} and \citet{Mosser} have notably determined the rotation rate in the core of red giants, and a rotation profile has even been obtained for the Sun \citep{Thompson} and a subgiant \citep{Deheuvels}.
Thanks to this promising technique, we are about to be given access both to the causes of the transport (rotation) and to its effects (chemical distribution).
We still need to extend its use to a larger number of stars, which is in progress, especially for CoRoT and Kepler stars. 

In many stellar evolution codes, rotationally induced mixing is taken into account thanks to a set of turbulent diffusion coefficients linked to various hydrodynamical instabilities triggered by differential rotation and meridional circulation.
Among these instabilities, shear instability is thought to produce the most efficient mixing \citep{KnoblochSpruit}, sometimes called ``shear mixing''.
This instability normally occurs in a stably stratified sheared flow when the Richardson number $Ri=(N/S)^2$, which compares stable stratification (with the Brunt-Väisälä frequency $N$) and the shear (with the shear rate $S={\rm d}U/{\rm d}z$), is lower than a critical value $Ri_{\rm c}$, which equals $1/4$ in the inviscid linear stability analysis \citep{Miles}.
However, except in some limited layers of evolved stars \citep{Hirschi}, the Richardson number is generally much larger than one in stars, and this criterion would result in almost always shear-stable stellar interiors.
In this high-Richardson-number or weak-shear regime, the very high thermal diffusivities found in stars are nevertheless able to reduce the amplitude of the buoyancy force through radiative losses and thus to overcome the stabilising effect of stratification \citep{Townsend, Zahn1974}.

We are interested here in one of the transport coefficients related to this instability, the radial diffusion coefficient due to radial differential rotation.
It was initially derived by \citet{Zahn1992} from the following phenomenological arguments.
\begin{itemize}
    \item Turbulent flows generally tend to reach a statistical steady state that is marginally stable with respect to the instability that has generated them.
    \item The destabilising effect of the very high thermal diffusivity $\kappa$ is such that the Richardson instability criterion should be replaced by $RiPe < Ri_{\rm c}$, where $Pe=UL/\kappa$ denotes the P\'eclet number based on velocity and length scales $U$ and $L$.
    \item The relevant P\'eclet number to use in this criterion is the turbulent one $Pe_\ell=u\ell/\kappa$, where $u$ is the velocity scale and $\ell$ the length scale of turbulence.
    \item The turbulent diffusion coefficient is proportional to the product of these turbulent scales: $D_{\rm t}\simeq u\ell/3$.
\end{itemize}
Zahn finally obtains a diffusion coefficient that reads
\begin{equation}
    D_{\rm t}\simeq\frac{\kappa}{3}\frac{Ri_{\rm c}}{Ri}.
\end{equation}
Since, several attemps have been made to add various physical ingredients to this model, such as $\mu$-gradients \citep{MaederMeynet1996, Maeder1997} and horizontal diffusion \citep{TalonZahn}.

As regards massive stars, the use of such models of rotational mixing in stellar evolution codes gives results closer to the observations than the standard model both for CNO products \citep{HegerLanger, MeynetMaeder2000, MaederMeynet2001} and for boron \citep{Fliegner, Mendel, Frischknecht}.
Thanks to the progress in the computational efficiency of such codes, it is now possible to compute entire synthetic stellar populations from a given distribution of initial conditions that include chemical composition and angular momentum.
As shown by \citet{Brott} and \citet{Potter}, these synthetic populations can be compared to observed ones, such as those obtained by large surveys.
In particular, the authors recover the main group of stars observed by \citet{Hunter}.

However, some observational constraints are still not explained by these models.
In particular, \citet{Hunter} are unable to explain the existence of nitrogen-enriched slow rotators which are actually observed.
Besides, current models do not predict enough transport of angular momentum to recover the seismic constraints on internal rotation in red giants and subgiants \citep{Eggenberger, Ceillier, Marques}.

Most of these phenomenological models of turbulent transport have never really been tested.
Indeed, the physical conditions present in stellar interiors are far too extreme to allow us to perform realistic laboratory experiments.
Our approach is to use direct numerical simulations (DNS) as experiments to test the existing prescriptions.
Local numerical simulations of homogeneous, sheared, and stably stratified turbulence (such as those performed in a geophysical context by \citealt{Gerz}, \citealt{Holt}, and \citealt{Jacobitz}) have the advantage of being as generic as possible while taking the essential ingredients (shear and stable stratification) into account.
There has already been an attempt to test Zahn's model by \citet{BruggenHillebrandt} but because their simulations used numerical viscosity and thermal diffusivity, which depend on the resolution, they were unable to study the effect of thermal diffusion.

In a previous paper \citep[hereafter Paper I]{PratLignieres}, we explored the dependence of the radial turbulent diffusion coefficient on thermal diffusity in the regime of the small P\'eclet numbers typical of stellar interiors.
This was done by considering a parallel-plane flow configuration with forced, uniform, vertical stable stratification and velocity shear.
Boundary conditions were periodic in the horizontal directions and impermeability, along with the mean shear, were imposed at the lower and upper boundaries.
Since the very high thermal diffusivity introduces a huge gap between the diffusive time scale and the dynamical one, which drastically increases the computational cost of complete Boussinesq simulations, we used an asymptotic development of the Boussinesq equations called the small-P\'eclet-number approximation (SPNA) described by \citet{Lignieres}.
We found that our simulations are qualitatively in good agreement with Zahn's prescription in this regime and we were able to give a quantitative estimate of the turbulent diffusion coefficient.

The purpose of the present paper is twofold.
On the one hand, we extend the previous study to the domain of large P\'eclet numbers, where Zahn's model is not valid any more.
Although the net effect on global mixing is often assumed to be negligible, some strong-shear situations exist, especially in evolved stars \citep[see for example][]{Hirschi}, where the Richardson number is small enough that stellar interiors are unstable with respect to the shear instability even on large scales, thus at large P\'eclet numbers.
On the other hand, we explore the dependence of the turbulent diffusion coefficient with respect to chemical stratification.
This effect is likely to be significant at the frontier of the radiative zone and a convective core.
Indeed, strong chemical gradients are expected in such a boundary, so that whether mixing occurs or not has a strong influence on stellar evolution.
Section~\ref{sec:form} presents the formalism and the methods used in our simulations.
Section~\ref{sec:large} is devoted to the case of a neutral chemical stratification, whereas Sect.~\ref{sec:chim} deals with the case of stable chemical stratification.
Then, the astrophysical consequences of our results are discussed in Sect.~\ref{sec:disc}, along with some prospects.

\section{Formalism and methods}
\label{sec:form}

The mathematical model and the numerical configuration used to perform the simulations presented in this paper are essentially the same as in Paper I, except for the dynamical effect of stable chemical stratification.
The flow is constituted by uniform, forced, vertical mean velocity shear and temperature and concentration gradients. 
In this configuration, turbulence can be homogeneous and stationary, which implies that turbulent transport of chemical elements can be described as a diffusive process.
In Sect.~\ref{sec:eq}, we recap the governing equations in the presence of this chemical stratification.
Then, we list in Sect.~\ref{sec:params} physical and numerical parameters on which our simulations depend.
Finally, Sect~\ref{sec:meth} is dedicated to the different methods of determining the turbulent diffusion coefficient that we have explored before choosing the current one.

\subsection{Governing equations}
\label{sec:eq}

The flow configuration, including boundary conditions and forcing terms, is detailed in Paper I so will not be reproduced here.
In contrast, the governing equations are modified to take the fluctuations of mean molecular weight due to concentration fluctuations into account.
The new equations are derived in Sect.~\ref{sec:bouss} from the Boussinesq equations, and the SPNA, which is valid for high thermal diffusivities, is then applied in Sect.~\ref{sec:spna}.

\subsubsection{Boussinesq approximation}
\label{sec:bouss}
In their classical form, the Boussinesq equations read
\begin{align}
    \nabla\cdot\vec v    &=  0,   \\
    \frac{\partial\vec v}{\partial t}+(\vec v\cdot\nabla)\vec v &=  -\frac{\nabla p}{\rho_0}+\frac{\rho'}{\rho_0}\vec g+\nu\Delta\vec v+\vec f_{\rm v},    \label{eq:momentum}
\end{align}
where $\vec v$ is the velocity of the fluid, $p$ the pressure deviation from hydrostatic equilibrium, $\rho_0$ the mean density, $\rho'$ the density fluctuations, $\vec g=-g\vec e_z$ the gravitational acceleration, $\nu$ the kinematic viscosity, and $\vec f_{\rm v}$ the forcing.
Density fluctuations can be expressed in terms of temperature and concentration (through mean molecular weight) fluctuations $\theta$ and $c'$ with respect to mean temperature and concentration profiles $T_0$ and $c_0$, here assumed to be linear:
\begin{equation}
    \frac{\rho'}{\rho}=-\alpha_{\rm T}\theta-\alpha_{\rm c}c',
\end{equation}
introducing the thermal expansion coefficient $\alpha_{\rm T}$ and its chemical equivalent $\alpha_{\rm c}$. 
Then, Eq.~\eqref{eq:momentum} becomes
\begin{equation}
    \frac{\partial\vec v}{\partial t}+(\vec v\cdot\nabla)\vec v = -\frac{\nabla p}{\rho_0}-(\alpha_{\rm T}\theta+\alpha_{\rm c}c')\vec g+\nu\Delta\vec v+\vec f_{\rm v},
\end{equation}
in which temperature $T=T_0+\theta$ and concentration $c=c_0+c'$ verify the advection/diffusion equations:
\begin{align}
    \frac{\partial\theta}{\partial t}+\vec v\cdot\nabla T   &=  \kappa\Delta\theta+f_{\rm T}, \\
    \frac{\partial c'}{\partial t}+\vec v\cdot\nabla c      &=  D_{\rm m}\Delta c'+f_{\rm c},
\end{align}
with $\kappa$ and $D_{\rm m}$ the thermal and molecular diffusivities, and $f_{\rm T}$ and $f_{\rm c}$ the thermal and chemical forcing terms.

The dimensionless form of the previous equations is obtained by using $L$, which is the characteristic size of the numerical domain, $S^{-1}$, $\Delta U=SL$, $\Delta T=L{\rm d}T_0/{\rm d}z$, $\Delta c=L{\rm d}c_0/{\rm d}z$, and $\rho_0S^2L^2$ as length, time, velocity, temperature, concentration, and pressure units, respectively.
Denoting dimensionless variables by a tilde, we write
\begin{align}
    \tilde\nabla\cdot\vec{\tilde v}   &=  0,    \\
    \frac{\partial\vec{\tilde v}}{\partial\tilde t}+(\vec{\tilde v}\cdot\tilde\nabla)\vec{\tilde v} &=  -\tilde\nabla\tilde p+\left(Ri\tilde\theta+Ri_{\mu}\tilde c'\right)\vec e_z+\frac{1}{Re}\tilde\Delta\vec{\tilde v}+\vec{\tilde f_{\rm v}}, \label{eq:mom}  \\
    \frac{\partial\tilde\theta}{\partial\tilde t}+\vec{\tilde v}\cdot\tilde\nabla\tilde\theta+\tilde v_z    &=  \frac{1}{Pe}\tilde\Delta\tilde\theta+\tilde f_{\rm T}, \label{eq:temp}\\
    \frac{\partial\tilde c'}{\partial\tilde t}+\vec{\tilde v}\cdot\tilde\nabla\tilde c'+\tilde v_z  &=  \frac{1}{Pe_{\rm c}}\tilde\Delta\tilde c'+\tilde f_{\rm c}, \label{eq:conc}
\end{align}
where there are five dimensionless parameters: the Richardson and P\'eclet numbers that we have already defined, the chemical Richardson number $Ri_\mu=\alpha_{\rm c}g/S^2{\rm d}c_0/{\rm d}z$, the Reynolds number $Re=SL^2/\nu$, and the chemical P\'eclet number $Pe_{\rm c}=SL^2/D_{\rm m}$.
In this paper, we study the dependence of the turbulent diffusion coefficient on the turbulent P\'eclet number and the chemical Richardson number.

\subsubsection{Small-P\'eclet number approximation}
\label{sec:spna}

Two conditions are required so that the stabilising effect of thermal stratification is affected by thermal diffusion.
Thermal diffusion and stratification must both play a dynamical role.
In other words, their characteristic time scales (the thermal diffusive one $\ell^2/\kappa$ and the buoyancy one $N^{-1}$) must be smaller than the dynamical one, which can be approximated by $S^{-1}$.
If we suppose that $u\simeq S\ell$, the first condition can be simply written $Pe_\ell < 1$.
The second one straightforwardly implies that $Ri > 1$.
This is the regime we want to study in stars, but when $Pe_\ell\ll1$, the problem becomes very costly to solve numerically without approximation, because the diffusive time scale becomes much smaller than the dynamical one.

Mathematically, this can be avoided by using the SPNA \citep{Lignieres}, which consists in a first-order Taylor expansion of the Boussinesq equations with respect to the P\'eclet number.
In this approximation, Eqs.~\eqref{eq:mom} and~\eqref{eq:temp} become
\begin{align}
    \frac{\partial\vec{\tilde v}}{\partial\tilde t}+(\vec{\tilde v}\cdot\tilde\nabla)\vec{\tilde v} &=  -\tilde\nabla\tilde p+\left(RiPe\tilde\psi+Ri_{\mu}\tilde c'\right)\vec e_z+\frac{1}{Re}\tilde\Delta\vec{\tilde v}+\vec{\tilde f_{\rm v}}, \\
    \tilde v_z                                                                                      &=  \tilde\Delta\tilde\psi,
\end{align}
where $\tilde\psi$ is defined by $\tilde\theta=Pe\tilde\psi$.
It can be noticed that the Richardson and P\'eclet numbers do not play independent roles in the SPNA.
Instead, the effects of both thermal stratification and diffusion become a single physical effect characterised by the number RiPe, which we call the ``Richardson-P\'eclet'' number, and its time scale is
\begin{equation}
    \label{eq:tau}
    \tau = \frac{\kappa}{N^2\ell^2}.
\end{equation}
In constrast, chemical stratification is not affected by thermal diffusion.

\subsection{Physical and numerical parameters of the simulations}
\label{sec:params}

In this study, we do not want to test Zahn's assumption that turbulent flows tend to reach a statistical steady state.
On the contrary, we tune the Richardson number (or the Richardson-P\'eclet number in the SPNA) to obtain this steady state.
We define the stationary Richarson number $Ri_{\rm s}$ as the highest Richardson number for which turbulent kinetic energy does not decrease statistically in time.
For $Ri<Ri_{\rm s}$, the flow eventually saturates into a statistical steady state, but at much larger scales, which we suspect to be close to the dimensions of the simulation domain.
We are not interested in such steady states because then the flow and the associated transport are expected to strongly depend on the size of the numerical domain.
We also define the stationary Richardson-P\'eclet number $(RiPe)_{\rm s}$ in the SPNA similarly.
In this regime, changing $RiPe$ is more or less equivalent to changing the turbulent scales, the flow evolving towards a steady state characterised by a unique $RiPe_\ell$.
For smaller $RiPe$, the turbulent scales are larger, and then get closer to the dimensions of the simulation domain.

Given the parameters we want to vary (the P\'eclet number and the chemical Richardson number) and those we want to be fixed (the Reynolds number and the chemical P\'eclet number), the only remaining physical parameters are thus linked to the initial conditions, which we discuss in Sect.~\ref{sec:init}.
They are also numerical parameters, namely the dimensions of the simulation domain (Sect.~\ref{sec:size}) and the resolution of our simulations (Sect.~\ref{sec:resol}).

\subsubsection{Initial conditions}
\label{sec:init}

Initial temperature and concentration fluctuations are set to zero.
For initial velocity fluctuations, we use an incompressible, isotropic field with a given power spectral density $E(k)$ generated thanks to the method explained by \citet{Orszag}.
Since the resulting velocity field is not a natural solution of the Navier-Stokes equations, it takes some time to reach a statistical steady state.
To reduce this relaxation time, \citet{Jacobitz} have deduced, from spectra of just relaxed velocity fields, the optimal power spectral density to use, that is,
\begin{equation}
    E(k) \propto k^2e^{-k^2/{k_0}^2},
\end{equation}
where $k$ is the norm of the wave vector and $k_0$ the value corresponding to the peak of the spectrum.
With this spectrum, initial conditions are now characterised by two parameters: the amplitude of initial fluctuations, which can be described by the quadratic mean of velocity fluctuations $q$, and the peak wave vector, which is linked to the horizontal integral length scale of turbulence $\ell_{\rm i}$, defined by
\begin{equation}
    \ell_{\rm i} = 2\pi\frac{\int_0^{+\infty}E(k_{\rm h})/k_{\rm h}{\rm d}k_{\rm h}}{\int_0^{+\infty}E(k_{\rm h}){\rm d}k_{\rm h}},
\end{equation}
where $k_{\rm h}$ is the norm of the horizontal wave vector.

\begin{figure}
    \resizebox{\hsize}{!}{\includegraphics{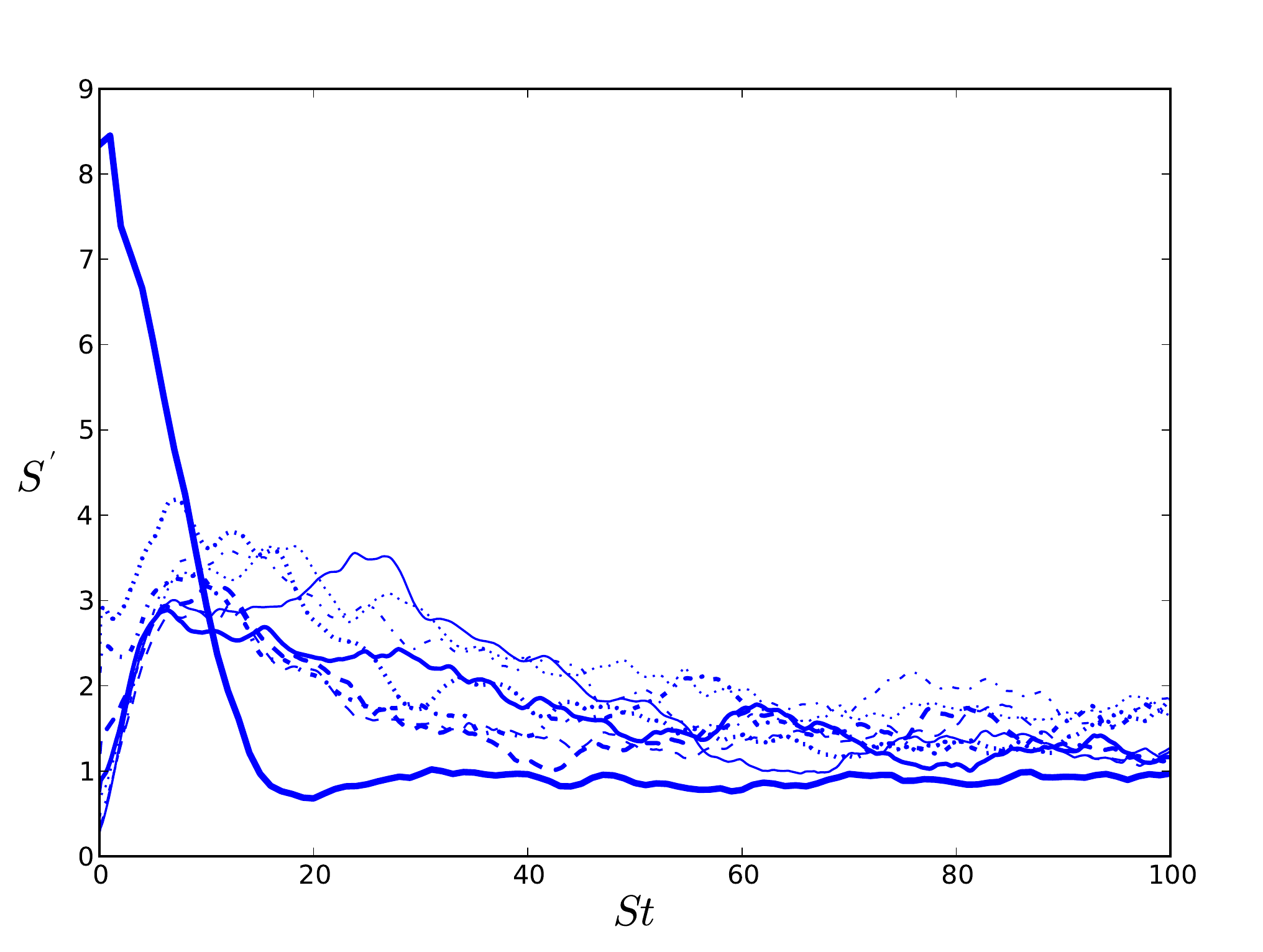}}
    \caption{Evolution of $S'$ as a function of time for different initial values}
    \label{fig:slq}
\end{figure}

From these parameters, we build two dimensionless parameters, the turbulent Reynolds number based on $q$ and $\ell_{\rm i}$ defined as
\begin{equation}
    Re_\ell = \frac{q\ell_{\rm i}}{\nu},
\end{equation}
and what we call the ``shear number'' $S'$ defined by
\begin{equation}
    S' = \frac{S\ell_{\rm i}}{q}.
\end{equation}
To determine the influence of initial conditions on the final statistical steady state, we have computed final values of these parameters as a function of their initial values.
The main result is that the final value of the shear number only slightly depends on its initial value, as illustrated in Fig.~\ref{fig:slq}.
For initial values ranging from 0.25 to 8.5, we obtain final values between 1 and 2.
Another comforting point about the generic nature of our simulations is that in those we have used in this paper, we observe a self-sustained mechanism similar to the one observed by \citet{Waleffe}.
This kind of mechanism, in which streamwise vortices transform into longitudinal streaks that disappear in favour of oscillating modes, which finally regenerate vortices, is known to be quite generic in shear flows.
In numerical simulations, it is present as long as the numerical domain and the Reynolds number are large enough.
Provided these conditions are met, the existence of the self-sustained process depends neither on the size of the simulation domain \citep{Yakhot} nor on the type of forcing \citep{Waleffe}.

\subsubsection{Dimensions of the numerical domain}
\label{sec:size}

Because of the stochastic nature of turbulence, the efficiency of turbulent transport is an averaged quantity, which can be obtained by statistical analyses.
It means that to obtain a physically relevant value of the transport coefficient, we need to have enough large structures (those which actually do the transport) in our simulation domain.
The more structures there are, the better the precision in determining the coefficient.
Nevertheless, increasing the number of large structures also leads to higher computational cost.

The size of large structures can be approximated by the integral scale $\ell_{\rm i}$.
For our simulations, we have chosen a numerical domain such that the vertical and horizontal dimensions $L_{\rm v}$ and $L_{\rm h}$ of the domain satisfy $L_{\rm h}/L_{\rm v}=\pi/4$.
In all the simulations, we have $0.13<\ell_{\rm i}/L_{\rm h}<0.21$ and $0.10<\ell_{\rm i}/L_{\rm v}<0.17$, which ensures that between four and eight large structures in each horizontal direction and at least six in the vertical direction are present in the numerical domain.

\subsubsection{Numerical resolution}
\label{sec:resol}
\begin{figure}
    \resizebox{\hsize}{!}{\includegraphics{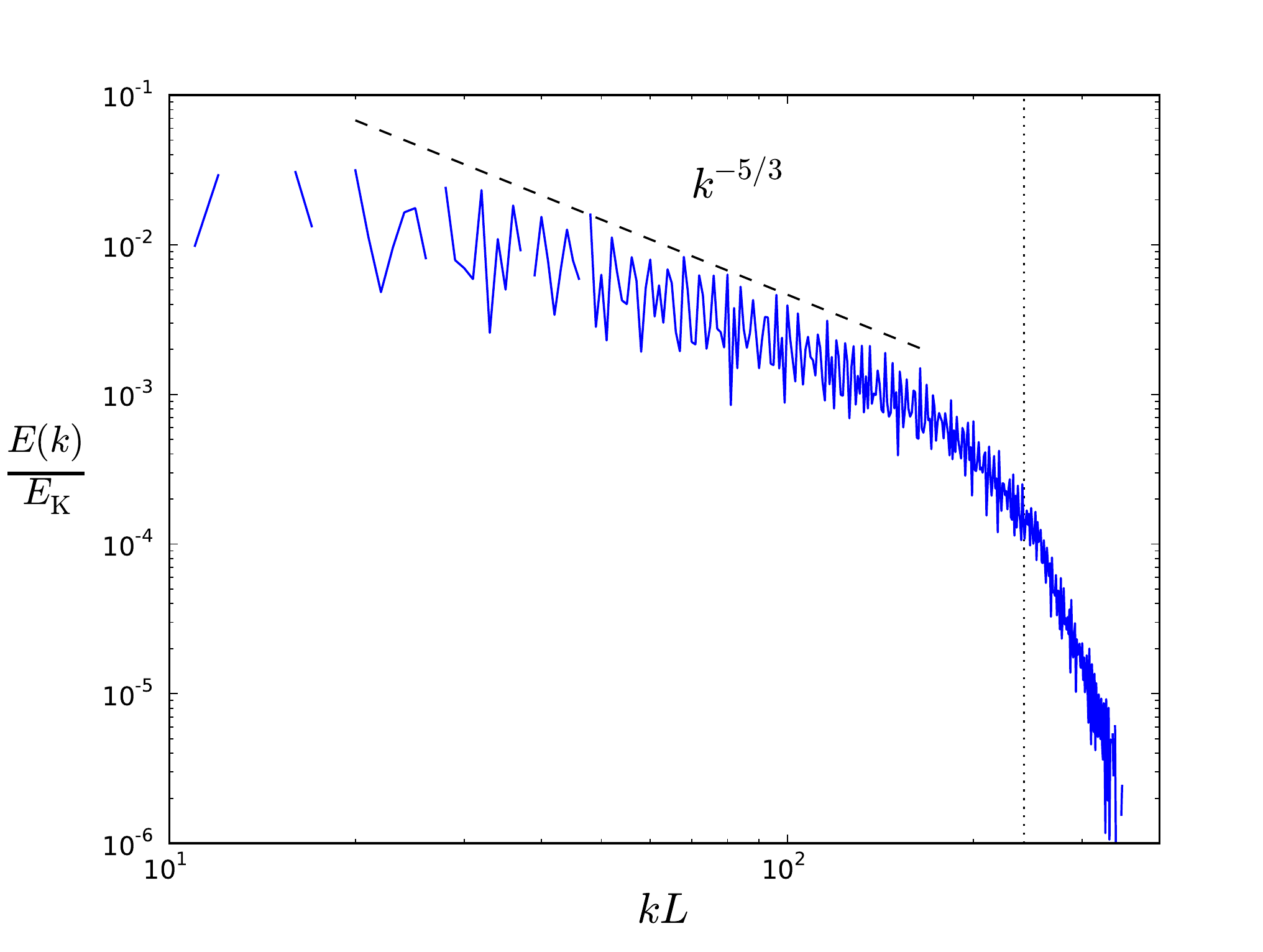}}
    \caption{Spectrum of turbulent kinetic energy. The dotted line corresponds to the largest relevant wave vector available in our simulations.}
    \label{fig:spec}
\end{figure}

Here, we consider DNS, where all scales down to the dissipation scale must be numerically resolved, as opposed to LES (large eddy simulations), where small scales are modelled with so-called sub-grid models.
Since the dissipation scale depends on the Reynolds number, the maximum value of the latter is limited by the resolution.
For the present configuration, 128 grid points in every horizontal direction and 257 ones in the vertical direction are enough to correctly resolve our simulations with a typical Reynolds number of $10^4$ (see Paper I for more details).
This is confirmed by the form of the spectrum of turbulent kinetic energy, presented in Fig.~\ref{fig:spec}.
One can clearly see that far away from the dissipation range, the spectrum follows a $k^{-5/3}$ law typical of Kolmogorov turbulence and rapidly drops once the dissipation scale is reached. 

\subsection{Test of the methods}
\label{sec:meth}

After the numerical setup is fixed, the transport coefficient can be determined by several methods.
We have tested three of them.
The first one is based on Lagrangian fluid particles (Sect.~\ref{sec:part}), the second one on the widening of a Gaussian mean concentration profile (Sect.~\ref{sec:gauss}), and the last one on a relation between the turbulent flux and the mean gradient of concentration (Sect. \ref{sec:flux}).

\subsubsection{Particle tracking}
\label{sec:part}

\begin{figure} 
    \resizebox{\hsize}{!}{\includegraphics{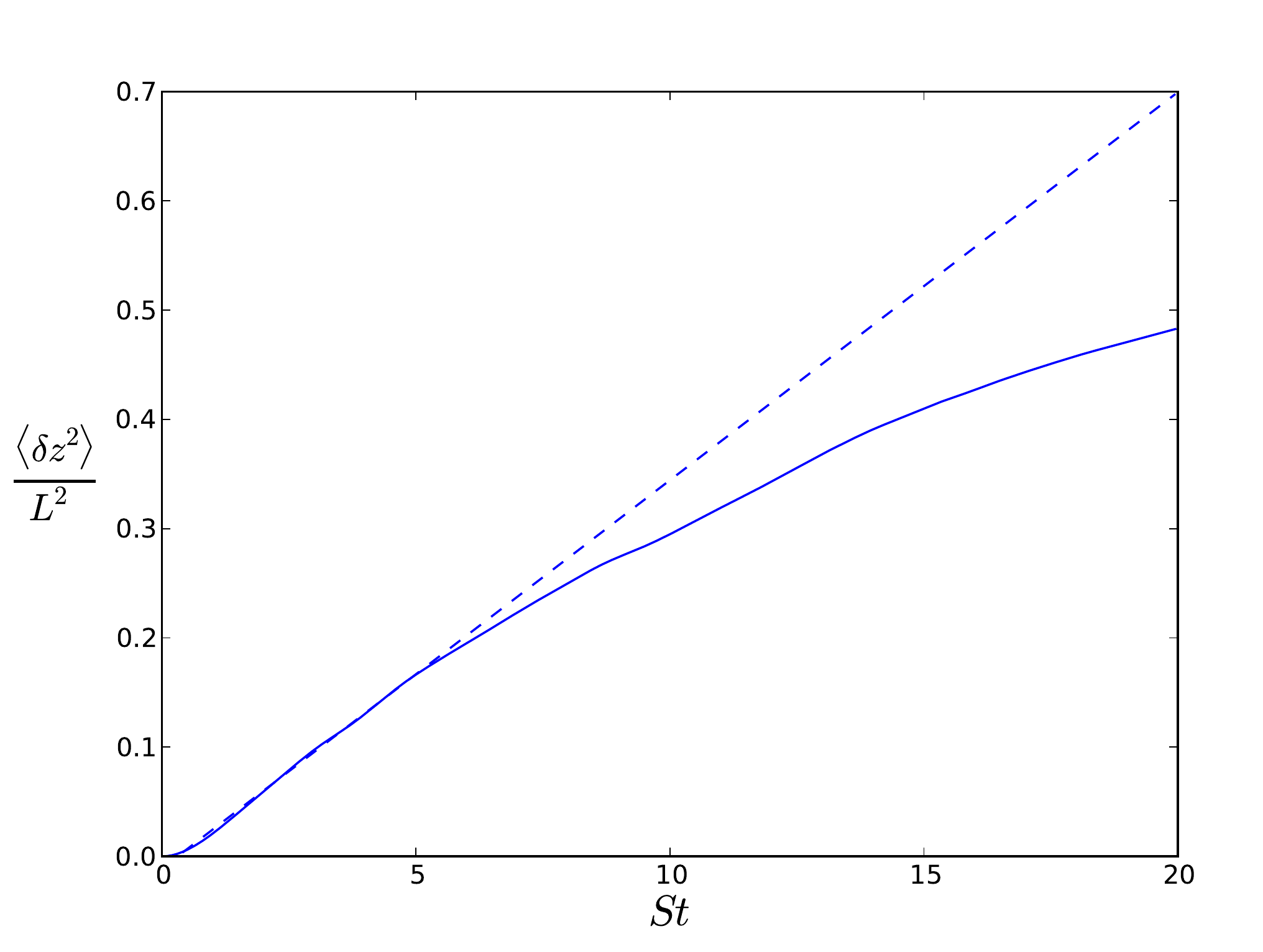}}
    \caption{Mean quadratic vertical dispersion against time.
        The solid line is the result of the simulation, and the dashed one represents the linear regression up to $St=5$.
        For $St<5$, the mean quadratic vertical dispersion grows linearly with time and saturates afterwards.}
    \label{fig:part}
\end{figure}

Since the considered flow is stationary and homogeneous from an Eulerian point of view away from the lower and upper boundaries (see Paper I for more details), our turbulence is homogeneous from a Lagrangian point of view in this region.
It means that the statistical properties of turbulence do not vary with time, seen from a fluid particle moving in the flow.
In these conditions, the theory of \citet{Taylor} applies and gives us the value of the transport coefficient $D_{\rm t}$ through the slope of the mean quadratic vertical dispersion of particles as a function of time:
\begin{equation}
    \langle\delta z^2\rangle=\langle[z(t)-z(0)]^2\rangle = 2D_{\rm t}t,
\end{equation}
where $\langle\rangle$ denotes an ensemble average or, equivalently, an average over particles.
Initially, we homogeneously put particles in the fluid and follow their displacements due to the velocity field.
Figure~\ref{fig:part} shows the evolution of the mean quadratic vertical dispersion with time for one of our simulations.
We observe that the transport is diffusive only for short times.

The non-diffusive regime appears when a significant part of particles have reached the lower or higher boundaries of the domain and thus cannot go further because of the impermeability of these boundaries.
Then, the turbulence is no longer homogeneous from a Lagrangian point of view.
As a consequence, the main fault of this method is that it only allows us to study turbulent transport for a limited time.
Because of natural fluctuations of the turbulent flow, determining the turbulent diffusion coefficient is thus subject to a large dispersion (up to 20\% of relative difference).

\subsubsection{Gaussian initial profile}
\label{sec:gauss}

Another method of estimating the transport coefficient is to study the evolution of the width of a Gaussian mean concentration field.
Whereas the previous method was a Lagrangian one, the present one is Eulerian in the sense that we solve Eq.~\eqref{eq:conc} (without the forcing term).
The mean vertical concentration profile is assumed to be a Gaussian given by
\begin{equation}
    C(z,t)=\frac{C_0}{\sigma(t)\sqrt{2\pi}}e^{-(z-z_0)^2/\left[2\sigma(t)^2\right]},
\end{equation}
where $z_0$ is the vertical position of the peak of the profile and $\sigma$ its standard deviation.
Indeed, if turbulent transport can be considered as a purely diffusive process, the profile remains Gaussian and the effective diffusion coefficient $D_{\rm eq}$ verifies
\begin{equation}
    \label{eq:sigma}
    D_{\rm eq} = \frac{1}{2}\frac{{\rm d}\sigma^2}{{\rm d}t}.
\end{equation}

\begin{figure}
    \resizebox{\hsize}{!}{\includegraphics{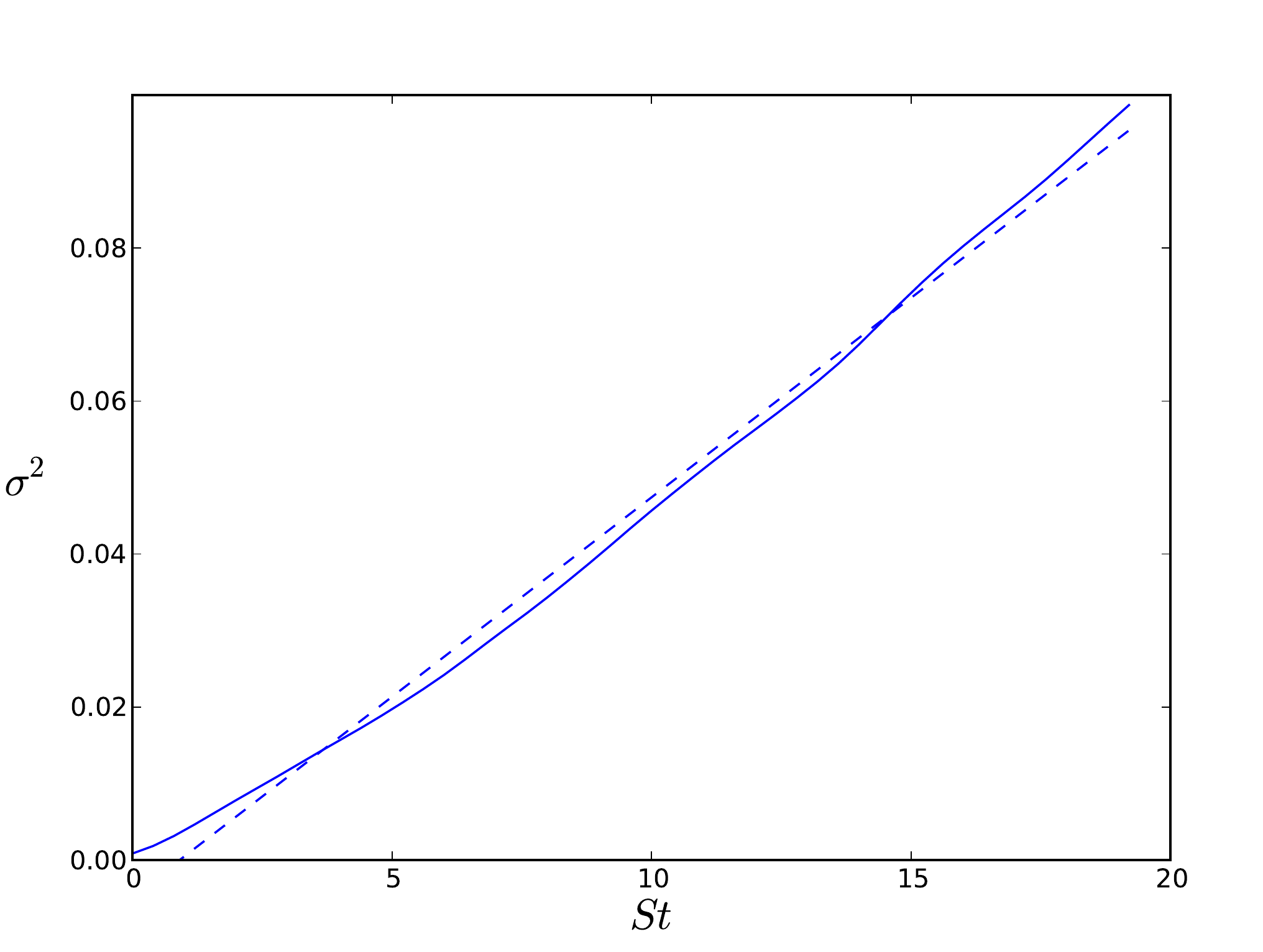}}
    \caption{Variance of the mean vertical concentration profile against time.
        The solid line corresponds to the simulation, and the dashed one to the affine regression.}
    \label{fig:sigma}
\end{figure}

Initially, the chemical element is mainly concentrated in the plane $z=z_0=L_{\rm v}/2$.
An example of evolution of $\sigma^2$ with time is represented in Fig.~\ref{fig:sigma}.
For short times, the variance of the mean profile grows linearly with time, thus indicating that the transport is diffusive, as already observed with the Lagrangian method.
Moreover, the values obtain by the two methods are similar (less than 15\% of relative difference).
For longer times, the Eulerian has a similar problem to the previous one: once the wings of the profile reach the boundaries, it is no longer Gaussian and Eq.~\eqref{eq:sigma} can no longer be used to estimate the transport coefficient.
Again, the limited averaging time of this method induces strong temporal fluctuations of the computed diffusion coefficient.

\subsubsection{Relation between turbulent flux and mean gradient}
\label{sec:flux}

In the presence of a mean concentration gradient, one can write the following relation between the mean flux of concentration fluctuations and the mean concentration gradient, provided that transport is diffusive, which is the case here:
\begin{equation}
    \langle c'v_z\rangle = -D_{\rm t}\frac{{\rm d}c_0}{{\rm d}z}.
\end{equation}
It is thus possible to determine the turbulent diffusion coefficient by measuring the mean concentration flux in the flow and computing
\begin{equation}
    D_{\rm t}=-\frac{\langle c'v_z\rangle}{{\rm d}c_0/{\rm d}z}.
\end{equation}
Thanks to the forcing term in Eq.~\eqref{eq:conc} combined with boundary conditions similar to those used for temperature, the mean gradient remains statistically constant, so that the transport coefficient can be averaged on much longer times than for the two other methods.
It follows that the precision on the estimate of this coefficient is better in this method, leading typically to relative fluctuations of only 5\% when averaging on several hundred shear times $S^{-1}$.
All the results published in Paper I and those presented thereafter have been obtained with this method. 

\section{Thermal diffusion in the chemically neutral case}
\label{sec:large}

This section deals with the effect of thermal diffusion on the vertical turbulent diffusion coefficient when chemical stratification is neutral ($Ri_\mu=0$).
To study this effect, we performed full Boussinesq simulations with $Re=10^4$ and $Pe$ ranging from 10 to $10^4$ (this corresponds to turbulent P\'eclet numbers $Pe_\ell=q\ell_{\rm i}/\kappa$, which are more relevant in this context, from 0.34 to 437), and one simulation in the small-P\'eclet-number approximation.
Table~\ref{tab:pel} displays the average values of some relevant physical parameters measured in the statistical steady state, including the turbulent Reynolds and P\'eclet numbers, the stationary Richardson number used to obtain the steady state, the product of the latter and the turbulent P\'eclet number, the ratio $\beta$ between the turbulent diffusion coefficient and the product of turbulent scales $u\ell_{\rm i}$, and the specific dissipation rate of thermal potential energy $\varepsilon_{\rm P}=\kappa\langle(\vec\nabla b)^2\rangle$, where $b=\alpha_{\rm T}g\theta/N$ is the ``buoyancy'', which is proportional to temperature fluctuations and has the dimension of a velocity.
\begin{table}[!ht]
    \caption{Simulation results for different turbulent P\'eclet numbers in the chemically neutral case}
    \label{tab:pel}
    \centering
    \(
    \begin{array}{ccccccc}
        \hline\hline
        Pe_\ell &   Re_\ell &   Ri_{\rm s}      &   Ri_{\rm s}Pe_\ell               &   \beta=D_{\rm t}/(q\ell_{\rm i}) &   \varepsilon_{\rm P}/(L^2S^{-3}) \\\hline
        \ll1    &   336     &   -               &   0.427\tablefootmark{$\star$}    &   0.133                   &   2.32\cdot10^{-4}  \\
        0.34    &   340     &   1.27            &   0.432                           &   0.139                   &   2.74\cdot10^{-4}  \\
        0.54    &   272     &   1.17            &   0.636                           &   0.110                   &   2.32\cdot10^{-4}  \\
        0.72    &   241     &   1.07            &   0.773                           &   9.29\cdot10^{-2}        &   2.16\cdot10^{-4}  \\
        0.90    &   150     &   1.10            &   0.990                           &   4.01\cdot10^{-2}        &   6.92\cdot10^{-5}  \\
        2.29    &   191     &   0.497           &   1.14                            &   6.23\cdot10^{-2}        &   1.50\cdot10^{-4}  \\
        5.62    &   225     &   0.250           &   1.41                            &   7.76\cdot10^{-2}        &   2.10\cdot10^{-4}  \\
        10.8    &   217     &   0.199           &   2.15                            &   7.03\cdot10^{-2}        &   1.98\cdot10^{-4}  \\
        24.4    &   244     &   0.15            &   3.65                            &   8.07\cdot10^{-2}        &   2.33\cdot10^{-4}  \\
        52      &   260     &   0.124           &   6.45                            &   0.104                   &   2.54\cdot10^{-4}  \\
        152     &   304     &   0.10            &   15.2                            &   0.109                   &   2.87\cdot10^{-4}  \\
        437     &   437     &   5.0\cdot10^{-2} &   21.9                            &   0.151                   &   2.94\cdot10^{-4}  \\
        \hline
    \end{array}
    \)
    \tablefoottext{$\star$}{The first simulation has been performed within the SPNA, and the number given in the $Ri_{\rm s}Pe_\ell$ column is actually $(RiPe_\ell)_{\rm s}$, which is defined as $(RiPe)_{\rm s}Re_\ell/Re$.}
\end{table}
Thermal potential energy is the potential energy linked to temperature fluctuations gained by a fluid particle moving away from equilibrium.

Firstly, we focus on the regime of small P\'eclet numbers, where new results confirm those presented in Paper~I (Sect.~\ref{sec:smallpe}).
Then, we consider the large-P\'eclet-number regime and compare our numerical results with existing models (Sect.~\ref{sec:largepe}).

\subsection{Small-P\'eclet-number regime}
\label{sec:smallpe}

In this section, we are interested in the regime of small P\'eclet numbers.
We performed numerical simulations both in the Boussinesq approximation at low Péclet number and in the SPNA.
We compare the results in Sect.~\ref{sec:res_smallpe} and try to interpret them in Sect.~\ref{sec:inter}.

\subsubsection{Results}
\label{sec:res_smallpe}

The first result is that Boussinesq simulations tend towards the SPNA one when the P\'eclet number decreases, thus validating the use of the SPNA in the non-linear regime for P\'eclet numbers smaller than 0.34.
Indeed, if we look at the results of the SPNA simulation and the Boussinesq one at $Pe_\ell=0.34$, we see that the value of $Ri_{\rm s}Pe_\ell$ in the Boussinesq simulation and that of $(RiPe_\ell)_{\rm s}$ in the SPNA one are very close to each other.
Moreover, the values of $\beta=D_{\rm t}/(q\ell_{\rm i})$ are also very similar in the two simulations.

\begin{figure}
    \resizebox{\hsize}{!}{\includegraphics{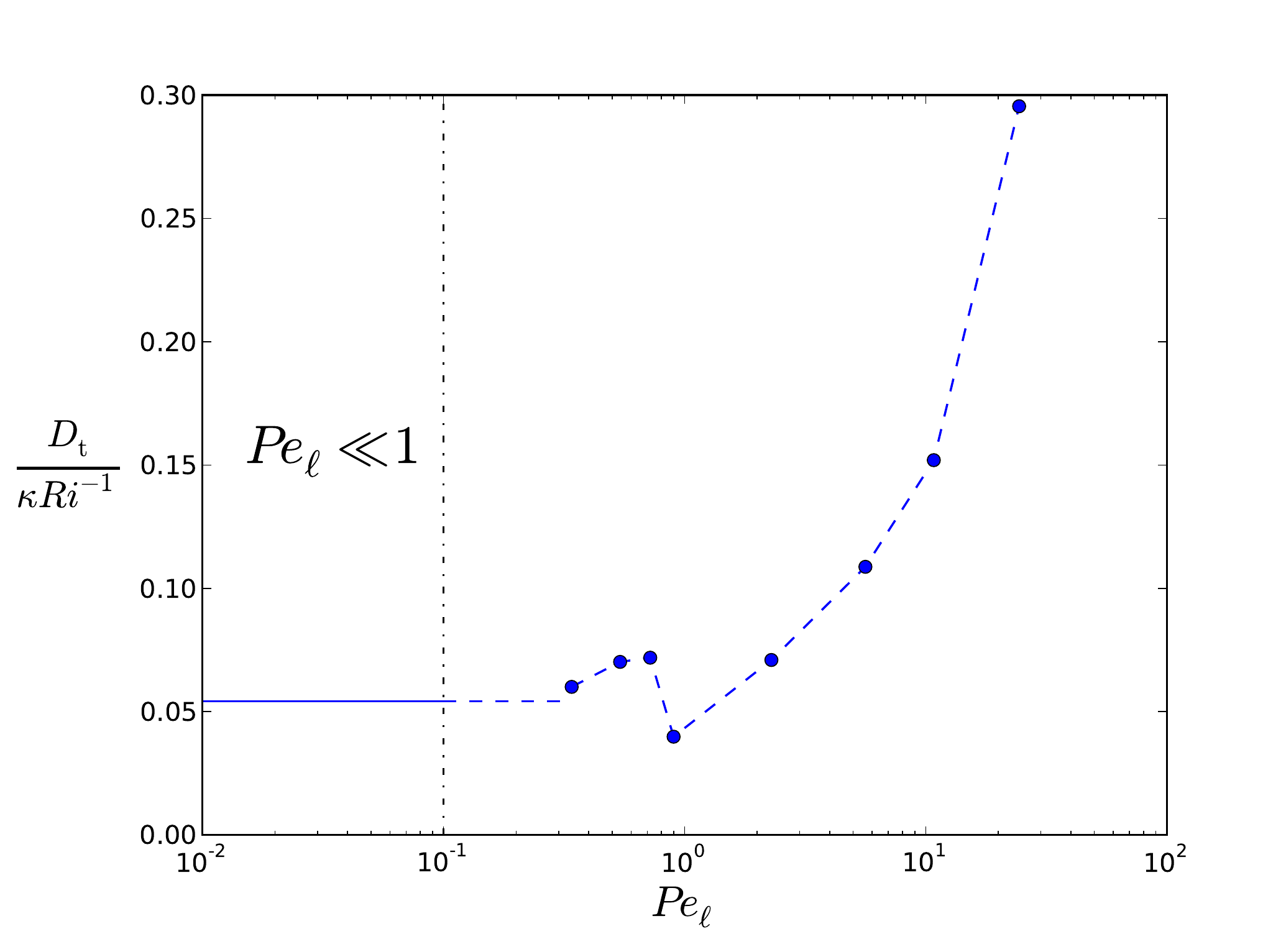}}
    \caption{$D_{\rm t}/(\kappa Ri^{-1})$ as a function of the turbulent P\'eclet number.
        Dots correspond to full Boussinesq simulations, whereas the solid line represents the value obtained in the SPNA one.}
    \label{fig:dtskappari}
\end{figure}

Furthermore, this proves that in this regime, stability is no longer characterised by the Richardson number alone, but by the product of the Richardson number and the turbulent P\'eclet number, as assumed by \cite{Zahn1974}.
The assumption that the turbulent diffusion coefficient is proportional to $u\ell_{\rm i}$ is also satisfied.
This seems to indicate that in the small-P\'eclet-number regime, the turbulent diffusion coefficient $D_{\rm t}$ is proportional to $\kappa/Ri$, and that this regime is already reached at $Pe_\ell=0.34$, which is not very low. 
This is confirmed by Fig.~\ref{fig:dtskappari}, which plots $D_{\rm t}/(\kappa Ri^{-1})$ against the turbulent P\'eclet number and suggests that for small P\'eclet numbers, $D_{\rm t}/(\kappa Ri^{-1})$ tends to a constant value that is close to what is obtained in the SPNA simulation. 

\subsubsection{Interpretation}
\label{sec:inter}

These results qualitatively agree with Zahn's model, but we also found a quantitative difference on the proportionality constant of about 30\%, as discussed in Paper I.
This is not surprising because Zahn's prescription gives only an order of magnitude for the diffusion coefficient.
One of the contributions of numerical simulations is precisely to give a quantitative estimate of such coefficients that cannot be deduced directly from theoretical arguments.

Another interesting physical discussion that was not made in Paper I is that thermal diffusion has two opposite effects on vertical motions: (i) decreasing the amplitude of the buoyancy force, which favours such motions, and (ii) contributing to the dissipation of kinetic energy, by dissipating potential energy, which inhibits these motions.
The first effect is taken into account in Zahn's model, but the second one, which is responsible for the damping of gravity modes, is not.
That despite this, Zahn's model seems to be valid can be understood by considering that when thermal diffusivity is very high, all available potential energy will be instantaneously dissipated, so that increasing the diffusivity does not dissipate more, whereas the amplitude of the buoyancy force is still decreasing.
The net effect is thus to reduce the stabilising effect of stratification, as accounted for by Zahn.
The balance between both effects may be an explanation for the strange behaviour observed in Fig.~\ref{fig:dtskappari} for $Pe_\ell=0.90$.

We propose now an alternative derivation of Zahn's model based on the mixing-length theory (MLT) in the SPNA.
The main idea is that turbulence can only occur if the dynamical time scale is smaller than that of the process which tends to inhibit it.
Here it is stable stratification modified by thermal diffusion, whose time scale $\tau$ is given in Eq.~\eqref{eq:tau}.
The previous condition reads as
\begin{equation}
    S^{-1} < \frac{\kappa}{N^2\ell^2},
\end{equation}
which yields
\begin{equation}
    \label{eq:l2inf}
    \ell^2 < \frac{\kappa S}{N^2}.
\end{equation}
We now assume that the largest scales allowed by this condition dominate the transport and that a mixing-length model $D_{\rm t}\sim S\ell^2$ holds. It follows that
\begin{equation}
    D_{\rm t} \sim \frac{\kappa}{Ri},
\end{equation}
which is the form of Zahn's model.
This derivation does not rely on the assumption of marginal stability or stationarity and may thus be applicable in cases where it is not satisfied.

\subsection{Large-P\'eclet-number regime}
\label{sec:largepe}

It is clear from Fig.~\ref{fig:dtskappari} that $D_{\rm t}/(\kappa Ri^{-1})={\rm ct}$ is not valid in the regime of large P\'eclet numbers.
\citet{Maeder1995} derived a prescription by following Zahn's approach but using a generalised Richardson criterion valid both for small and large P\'eclet numbers.
This problem has also been studied in geophysics ($Pr=\nu/\kappa=Pe/Re\sim 1$ in the atmosphere or the oceans), notably by \citet{LindborgBrethouwer}.
Section~\ref{sec:maeder95}, we compare our simulations, already presented in Table.~\ref{tab:pel}, to Maeder's model and show they are not compatible.
Then, we explain how the model proposed by \citet{LindborgBrethouwer} is supported by our simulations in Sect.~\ref{sec:lindborg}.

\subsubsection{Comparison with Maeder's model}
\label{sec:maeder95}

\begin{figure}
    \resizebox{\hsize}{!}{\includegraphics{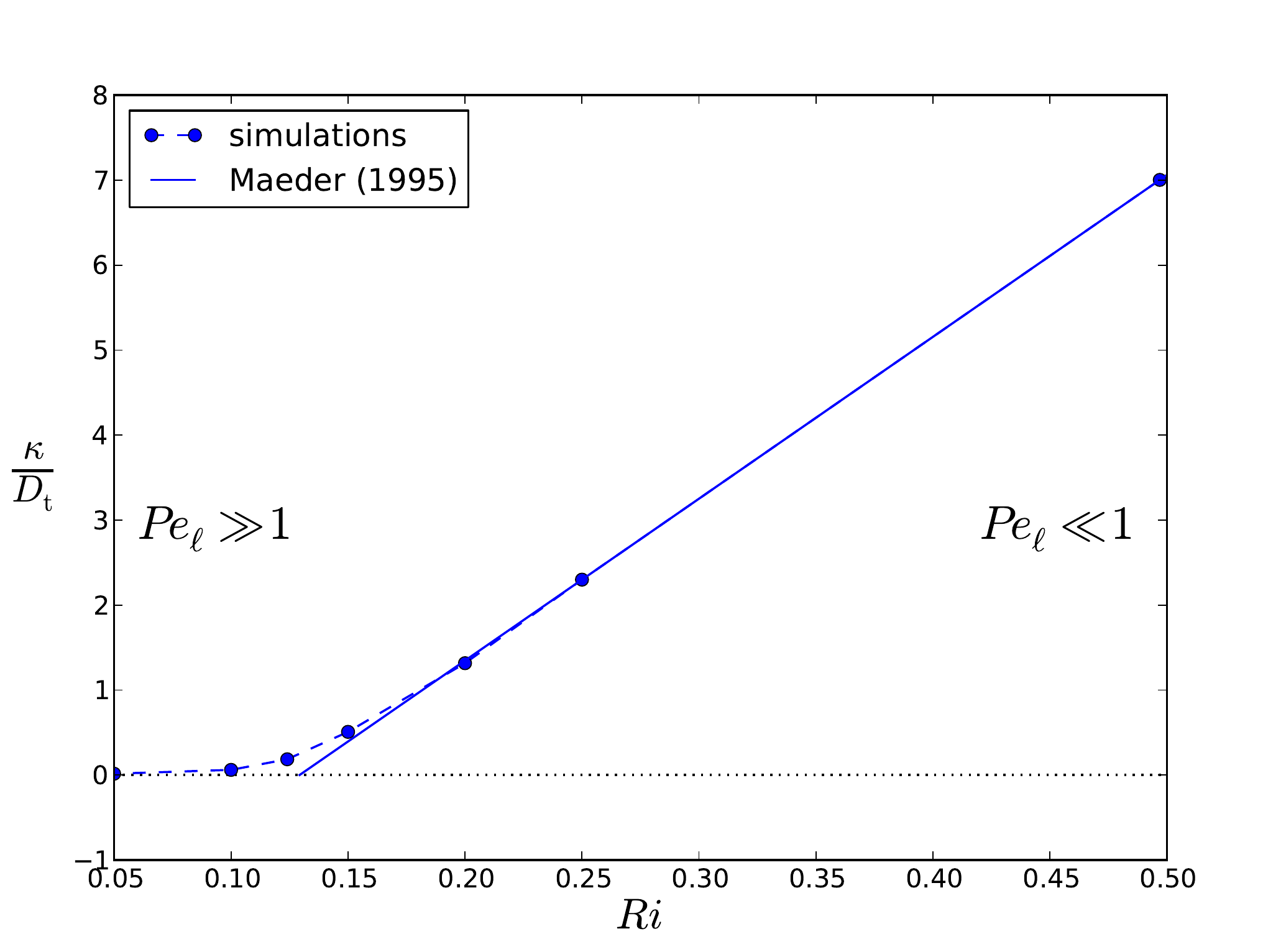}}
    \caption{$\kappa/D_{\rm t}$ against the Richardson number.
        Dots correspond to our simulations, whereas the solid line represents Maeder's model calibrated in the small-P\'eclet-number regime, where it reduces to Zahn's model.}
    \label{fig:maeder}
\end{figure}

The idea of \citet{Maeder1995} is to use a generalised Richardson criterion for marginal stability of the form
\begin{equation}
    \label{eq:maeder_crit}
    Ri\frac{Pe_\ell}{1+Pe_\ell} = Ri_{\rm c},
\end{equation}
which reduces to the classical one $Ri=Ri_{\rm c}$ for large P\'eclet numbers and to that of \citet{Zahn1974} for small ones.
This criterion then leads to the following expression for the turbulent diffusion coefficient:
\begin{equation}
    D_{\rm t}=\beta\kappa\frac{Ri_{\rm c}}{Ri - Ri_{\rm c}}.
\end{equation}
To compare it with our simulations, the previous relation can be written
\begin{equation}
    \frac{\kappa}{D_{\rm t}} = \frac{Ri-Ri_{\rm c}}{\beta Ri_{\rm c}},
\end{equation}
which shows that in Maeder's model there is an affine relation between $\kappa/D_{\rm t}$ and the Richardson number. 
As illustrated in Fig.~\ref{fig:maeder}, such a relation is not supported by our simulations for $Ri<0.15$, which corresponds to $Pe_\ell>25$.
Indeed, Maeder's model predicts that at large P\'eclet numbers, the stationary Richardson number should tend towards $Ri_{\rm c}$, which is supposed to be constant, whereas in our simulations $Ri_{\rm s}$ reaches lower values.

Maeder's model was intended to be a general model valid for all P\'eclet numbers.
If this model agrees with our numerical results in the small-P\'eclet-number regime, where it reduces to Zahn's, this is no longer the case in the large-P\'eclet-number regime, which is however the regime it has been created for.
One potential reason is that Eq.~\eqref{eq:maeder_crit} assumes that thermal diffusion modifies marginal stability even at a high P\'eclet number, a property that is not found in the linear stability analysis \citep{LignieresEtAl}.
As we see next, thermal diffusion does play an important role in radial transport in the high-Péclet-number regime, but only on motions with small enough length scales, which are produced by the turbulent cascade from the instability injection scale.

\subsubsection{Comparison with Lindborg's model}
\label{sec:lindborg}

\begin{figure}
    \resizebox{\hsize}{!}{\includegraphics{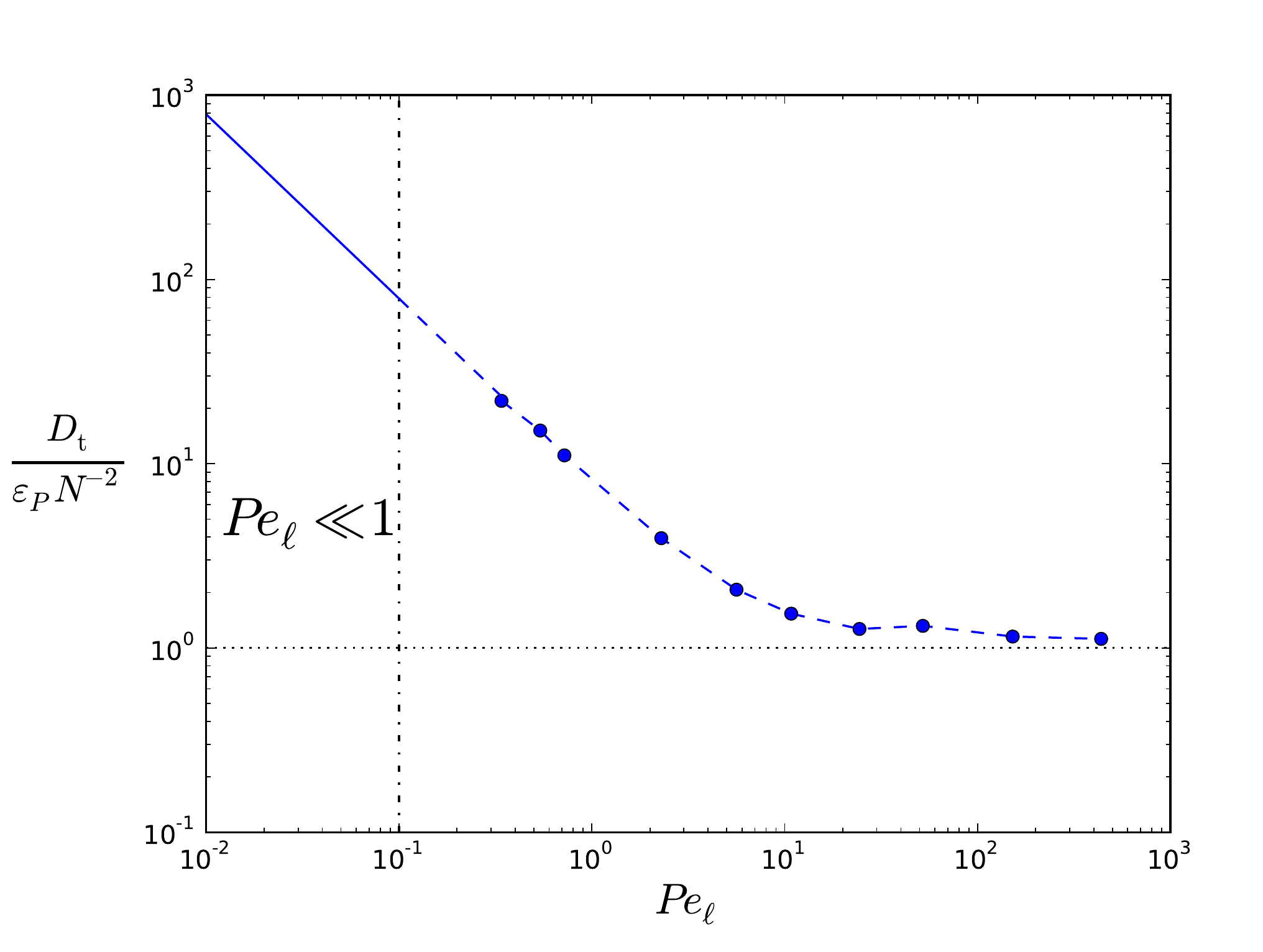}}
    \caption{$D_{\rm t}/(\varepsilon_{\rm P}N^{-2})$ against the turbulent P\'eclet number.
        Same legend as in Fig.~\ref{fig:dtskappari}.}
    \label{fig:lindborg}
\end{figure}

Lindborg's model is based on the buoyancy equation of a Lagrangian fluid particle that reads
\begin{equation}
    \frac{{\rm d}b}{{\rm d}t}+Nv_z = \kappa\Delta b.
\end{equation}
This equation can be used to describe the temporal evolution of the mean vertical dispersion of fluid particles.
Thanks to certain assumptions on the properties of turbulence, including Lagrangian statistical homogeneity, \citet{LindborgBrethouwer} are thus able to write
\begin{equation}
    \langle \delta z^2\rangle=\frac{\langle\delta b^2\rangle}{N^2}+2\frac{\varepsilon_{\rm P}}{N^2}t,
\end{equation}
where $\langle\delta b^2\rangle$ is the mean buoyancy dispersion and $\varepsilon_{\rm P}$ the specific dissipation rate of potential energy linked to buoyancy.
At late times, the first term, which corresponds to the case of zero thermal diffusion, tends to a constant value, whereas the second one increases linearly with time.
The corresponding diffusion coefficient is
\begin{equation}
    D_{\rm t}=\frac{\varepsilon_{\rm P}}{N^2}.
\end{equation}
One interesting property of this model is that it does not involve any free parameter, which is not the case for many phenomenological models currently used in stellar evolution codes.
We tested this model by computing the ratio $D_{\rm t}/(\varepsilon_{\rm P}N^{-2})$ for the different simulations of Table~\ref{tab:pel}.
As illustrated in Fig.~\ref{fig:lindborg}, one notices that $D_{\rm t}/(\varepsilon_{\rm P}N^{-2})$ tends towards one for large P\'eclet numbers.

\begin{figure}
    \resizebox{\hsize}{!}{\includegraphics{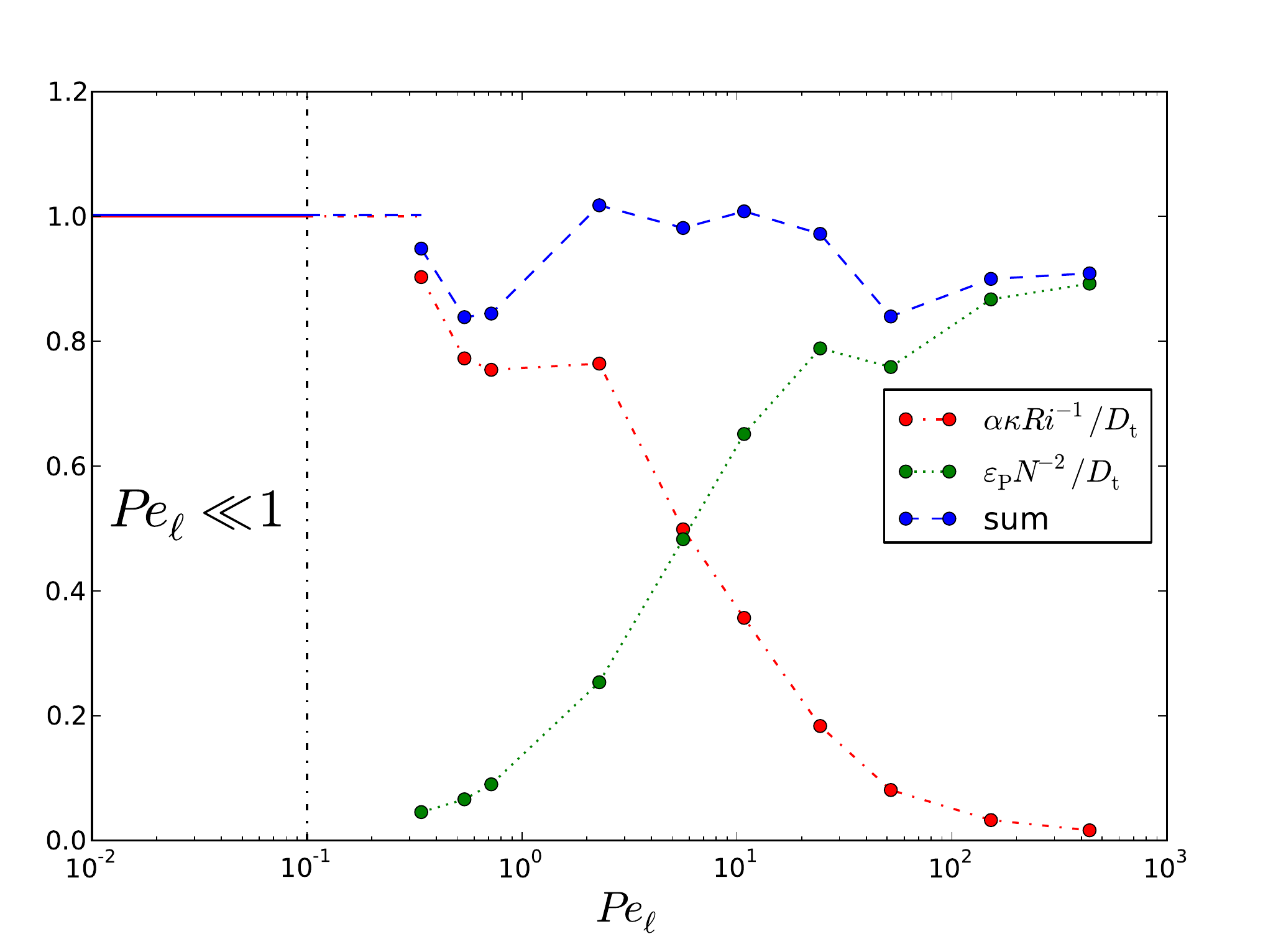}}
    \caption{Ratios between Zahn's and Lindborg's models and the total transport observed in our simulations.
        Same legend as in Fig.~\ref{fig:dtskappari}.}
    \label{fig:sum}
\end{figure}

Lindborg's model is thus valid for large P\'eclet numbers, but clearly not for small ones.
This is because \citet{LindborgBrethouwer} neglected in their study a term that is the dominant one for high thermal diffusivities.
Based on this observation, one can then wonder if it is possible to derive a general model based on Zahn's and Lindborg's models, which are each valid in its own regime, by taking the sum.
Figure~\ref{fig:sum} thus compares the ratios between each model and the total transport observed in our simulations.
The important result is that the ratio between the sum of the two models and the total observed transport is generally close to one.
We can thus write
\begin{equation}
    D_{\rm t} = \beta\kappa\frac{Ri_{\rm c}}{Ri}+\frac{\varepsilon_{\rm P}}{N^2} = \frac{1}{Ri}\left(\beta\kappa Ri_{\rm c} + \frac{\varepsilon_{\rm P}}{S^2}\right),
\end{equation}
which is approximately valid for all P\'eclet numbers.
This prescription for turbulent transport cannot be used in stellar evolution codes as long as $\varepsilon_{\rm P}$ is not expressed in terms of the mean quantities characterising the flow in these codes.
This will be done in a future work.

\section{Chemical stratification}
\label{sec:chim}

In this section we are interested in the dependence of the turbulent diffusion coefficient with respect to chemical stratification.
To study this dependence, we performed several simulations with different values of the chemical Richardson number.
These simulations and the empirical form of the turbulent diffusion coefficient deduced from them are presented in Sect.~\ref{sec:res_rimu}.
In Sect.~\ref{sec:comp_rimu}, we compare the form of the diffusion coefficient with existing prescriptions currently used in stellar evolution codes.
The results are discussed in Sect.~\ref{sec:inter_rimu}.

\subsection{Results}
\label{sec:res_rimu}

\begin{figure}
    \resizebox{\hsize}{!}{\includegraphics{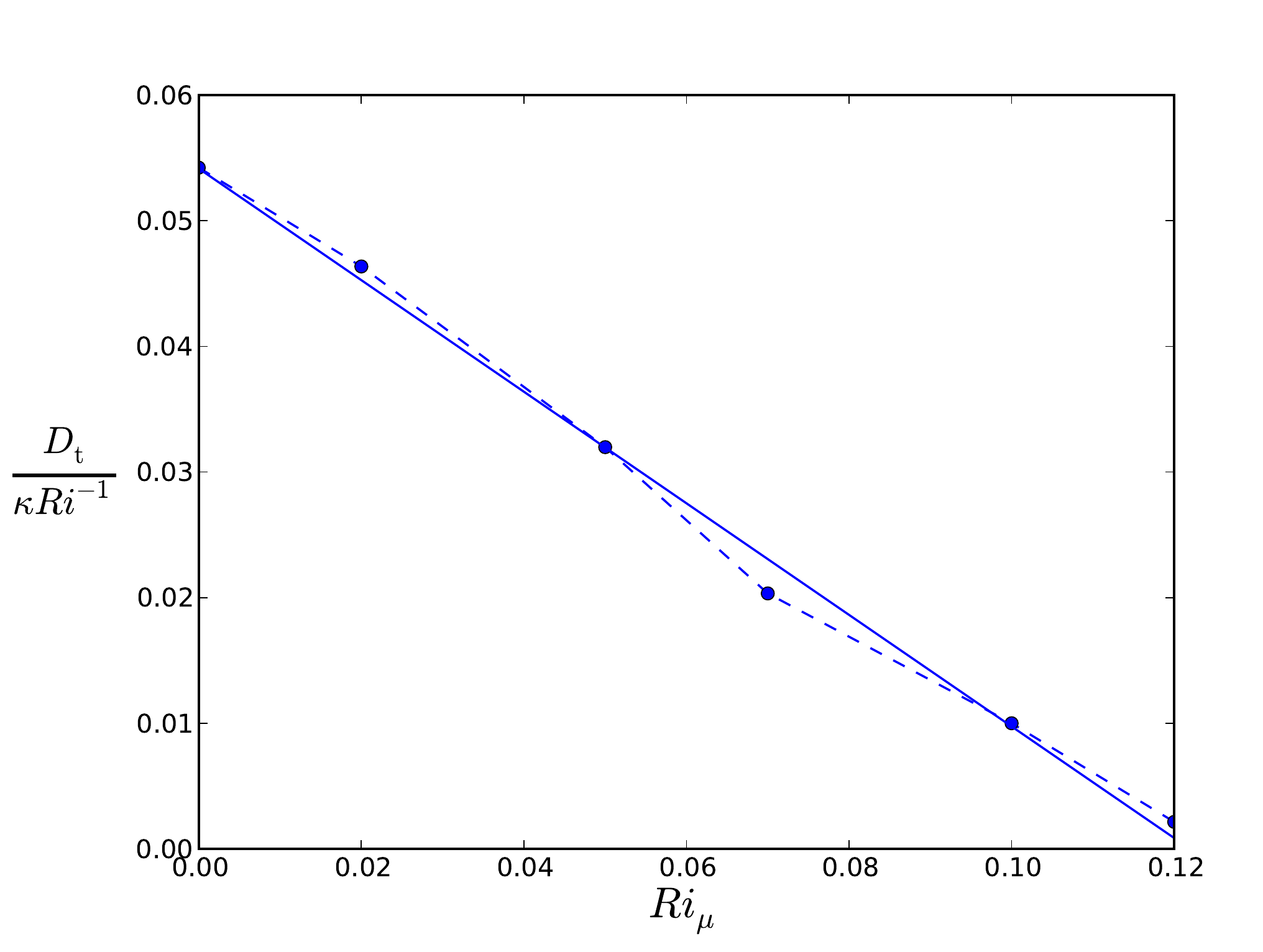}}
    \caption{$D_{\rm t}/(\kappa Ri^{-1})$ against the chemical Richardson number.
        Dots correspond to our simulations and the solid line to the affine regression.}
    \label{fig:rimu}
\end{figure}

Since the SPNA gives satisfying results in the neutral case, we have chosen to use it here, too.
For each value of the chemical Richardson number, we have tuned the Richardson-P\'eclet number to obtain a statistical steady state and determined the turbulent diffusion coefficient in this state. 
The results of our simulations are summarised in Table~\ref{tab:rimu}.
\begin{table}
    \caption{Simulation results for different chemical Richardson numbers}
    \label{tab:rimu}
    \centering
    \(
    \begin{array}{ccccc}
        \hline\hline
        Ri_\mu  &   (RiPe)_{\rm s}          &   (RiPe_\ell)_{\rm s}         &   Re_\ell &   D_{\rm t}/(SL^2)    \\\hline
        0       &   12.7                    &   0.427                       &   336     &   4.27\cdot10^{-3}    \\
        0.02    &   13.4                    &   0.393                       &   293     &   3.46\cdot10^{-3}    \\
        0.05    &   13.0                    &   0.335                       &   258     &   2.46\cdot10^{-3}  \\
        0.07    &   8.2                     &   0.219                       &   267     &   2.48\cdot10^{-3}  \\
        0.1     &   5.0                     &   0.123                       &   245     &   2.00\cdot10^{-3}  \\
        0.12    &   1.0                     &   2.59\cdot10^{-2}            &   260     &   2.15\cdot10^{-3}  \\\hline
    \end{array}
    \)
\end{table}
It is important to note that no stationary state could be reached for chemical Richardson numbers greater than 0.12, since turbulence decays regardless of the value of the thermal Richardson number.
Figure~\ref{fig:rimu} displays $D_{\rm t}/(\kappa Ri^{-1})$ as a function of the chemical Richardson number.
It appears that this plot is described well by an affine relation that is written
\begin{equation}
    \frac{D_{\rm t}}{\kappa Ri^{-1}} = \beta(Ri_{\rm cr}-Ri_\mu),
\end{equation}
with $\beta=0.45$ and $Ri_{\rm cr}=0.12$.
The turbulent diffusion coefficient is then given by
\begin{equation}
    \label{eq:dtrimu}
    D_{\rm t} = \beta\kappa\frac{Ri_{\rm cr}-Ri_\mu}{Ri}.
\end{equation}
Unlike the neutral case in the small-P\'eclet-number regime, where we found that $D_{\rm t}$ only depends on one unknown constant, namely the product $\beta Ri_{\rm c} = D_{\rm t}Ri/\kappa$, here it depends on two constants $\beta$ and $Ri_{\rm cr}$ that can be determined independently and that indeed play different physical roles.

\subsection{Comparison with existing models}
\label{sec:comp_rimu}

Several models of turbulent transport in the presence of a mean gradient of mean molecular weight have been proposed for use in stellar evolution codes, and we have compared our simulations with two of them: \citet{MaederMeynet1996} and \citet{Maeder1997}.
In particular, we present in Sect.~\ref{sec:maeder_meynet} to what extent our prescription agrees with the model of \citet{MaederMeynet1996}.
Then we explain in Sect.~\ref{sec:maeder97} why our simulations are incompatible with the model of \citet{Maeder1997}.

\subsubsection{Maeder \& Meynet's model}
\label{sec:maeder_meynet}

\citet{MaederMeynet1996} extended the model of \citet{Maeder1995} to the chemically stratified case.
As seen before, we do not expect such a model to be valid for large P\'eclet numbers.
However, for small ones, the authors found a form of the turbulent diffusion coefficient that is similar to Eq.~\eqref{eq:dtrimu}.
To obtain this result, they assumed that the relevant marginal stability criterion reads as
\begin{equation}
    \label{eq:crit_mm}
    RiPe_\ell+Ri_\mu = Ri_{\rm c},
\end{equation}
where $Ri_{\rm c}$ can be seen as a critical chemical Richardson number.
Assuming further that $D_{\rm t}=\beta u\ell$, one obtains the same form as Eq.~\eqref{eq:dtrimu}.
Since the authors used the same numerical values for $\beta$ and $Ri_{\rm c}$ as Zahn, 1/3 and 1/4, respectively, their model shows quantitative differences with our numerical results: around 25\% of relative difference for $\beta$ and 50\% for $Ri_{\rm c}$.
An important aspect of this criterion is that chemical stratification is not affected by thermal diffusion.
Thus, a sufficiently stable chemical stratification is able to totally inhibit turbulent transport.

\subsubsection{Maeder's model}
\label{sec:maeder97}

In contrast, the model of \citet{Maeder1997} is based on the assumption that mixing can occur for any value of the chemical Richardson number, provided that the flow is thermally unstable.
As a consequence, stability is then given by the criterion of \citet{Zahn1974}, and the turbulent diffusion coefficient is
\begin{equation}
    D_{\rm t}=\beta\kappa\frac{Ri_{\rm c}}{Ri+Ri_\mu}.
\end{equation}
One can easily notice that when $\kappa$ and $Ri$ are very high and $Ri_\mu$ remains finite, this coefficient reduces to that of \cite{Zahn1992}, which does not depend on the chemical Richardson number.
This is clearly not what we observe in our simulations.
Maeder's model is thus incompatible with our results.

\subsection{Interpretation}
\label{sec:inter_rimu}

Looking at the previous results, one may be tempted to conclude that the model of \citet{MaederMeynet1996} is more realistic than that of \citet{Maeder1997} and therefore should be used in stellar evolution codes.
However, \citet{Meynet} find that \citet{Maeder1997} is the prescription that has the best fit to the observation.
This is likely to be because this prescription enables mixing in strongly chemically stratified zones, such as the frontier between the convective core and the radiative envelope of massive stars.
The model of \citet{MaederMeynet1996} was not included in this study, but it predicts a less efficient mixing in such zones than those considered in the study.
If we admit that \citet{Maeder1997} overestimates the mixing efficiency in radiative zones of stars, then an additional source of mixing is needed to explain why it obtains such good agreement with the observations.

\section{Discussion}
\label{sec:disc}

We have tested several models of turbulent transport.
For large P\'eclet numbers, our simulations invalidate the model of \citet{Maeder1995} and validate that of \citet{LindborgBrethouwer}.
Nevertheless, the latter cannot be applied as such in stellar evolution codes.
As regards the dynamical effect of chemical stratification, our simulations agree with a special case of the model of \citet{MaederMeynet1996}, but not with that of \citet{Maeder1997}, which nevertheless gives the better results when compared with observations.

These results were obtained with a flow configuration that depends on several physical and numerical parameters, such as initial conditions and the Reynolds number.
As regards initial conditions, we have chosen a parameter domain in which the flow is likely to be the most generic, as discussed in Sect.~\ref{sec:init}.
For the Reynolds number, a preliminary study shows that transport does not significantly vary for turbulent Reynolds numbers ranging from 225 to 340.
According to \citet{MichaudZahn}, these values are consistent with what we expect in stellar radiative zones. 

In addition, this configuration relies on certain hypotheses, including stationarity and homogeneity, that are forced in an uncommon way.
This configuration is generic, but not necessarily natural.
Other numerical methods such as shearing boxes, used notably in geophysics \citep[e.g.][]{Jacobitz} and in astrophysics for the study of accretion discs \citep[e.g.][]{Lesur}, also force uniform mean shear flows, but not exactly in the same way as in the present paper.
It would be interesting to see if our results can be recovered with such approaches.
Moreover, unstationary and unhomogeneous configurations should also be considered.
As an example, direct numerical simulations of free shear layers, such as in \citet{BruggenHillebrandt}, could be considered.

There are still many other physical ingredients that could play a role in the radial transport due to shear instability.
Among them is the effect of a very efficient horizontal turbulent diffusion, which is said to reduce the stabilising effect of chemical stratification \citep{TalonZahn}.
One can also wonder what the effects of rotation (by introducing the Coriolis force in the simulations) and magnetic field are on shear instability.
An attempt has been made recently by \citet{Maeder2013} to provide some general diffusion coefficients, when simultaneously taking several hydrodynamical instabilities into account.
As regards magnetic field, \citet{Lecoanet} show that in the linear regime particular magnetic configurations can destabilise a flow, which would be stable without it.
Whether this result remains relevant in the non-linear regime is still an open question.

Until now, we have computed transport coefficients for chemical elements, but turbulent viscosity is also crucial to understanding angular momentum transport and should be estimated in the same way.
In a forthcoming paper, we intend to determine relations between turbulent viscosity and turbulent thermal and chemical diffusivities.

\begin{acknowledgements}
    We thank Georges Meynet for his useful comments.
    This work was granted access to the HPC resources of CALMIP under the allocation 2013--P0507.
\end{acknowledgements}

\bibliographystyle{aa}
\bibliography{refs}

\end{document}